\documentclass[english,preprint]{aastex}
\usepackage{graphics}

\newcommand{\kms}{km~s\ensuremath{^{-1}}}
\newcommand{\pcs}{photons cm\ensuremath{^{-2}} s\ensuremath{^{-1}}}
\newcommand{\vlsr}{\ensuremath{v_{\rm LSR}}}
\newcommand{\hi}{\ion{H}{1}} 
\newcommand{\sii}{\ion{S}{2}} \newcommand{\nii}{\ion{N}{2}}
\newcommand{\ha}{H\ensuremath{\alpha}}

\begin{document}

\title{A Map of the Ionized Component \\
  of the Intermediate Velocity Cloud Complex K}

\author{L. M. Haffner, R. J. Reynolds}

\affil{Department of Astronomy, University of Wisconsin, Madison, WI 53706}

\email{haffner@astro.wisc.edu, reynolds@astro.wisc.edu}

\and

\author{S. L. Tufte}

\affil{Department of Physics, Lewis and Clark College, Portland, OR 97219}

\email{tufte@lclarke.edu}

\begin{abstract}
  The Wisconsin H-Alpha Mapper (WHAM) Northern Sky Survey is revealing
  that many intermediate-velocity (\( |\vlsr |\leq 100 \) \kms)
  neutral clouds and complexes have an associated ionized component.
  We present the first map of the \ha\ emission from an intermediate-
  or high-velocity structure: Complex~K. This large, high-latitude
  feature stretches from \( \ell \approx 10\arcdeg \) to \( 70\arcdeg
  \), \( b\approx +30\arcdeg \) to \( +70\arcdeg \) and peaks in
  velocity over \( \vlsr \approx -60 \) to \( -80 \) \kms. The neutral
  and ionized gas generally trace each other quite well in the
  complex, but the detailed structure is not identical. In particular,
  the \ha\ emission peaks in brightness at slightly higher Galactic
  longitudes than corresponding 21 cm features.  The ionized gas has a
  peak \ha\ intensity of 0.5 Rayleighs, corresponding to an emission
  measure of 1.1 cm\( ^{-6} \) pc. Structures in the complex are
  traced by WHAM down to about 0.1 Rayleighs (0.2 cm\( ^{-6} \) pc).
  Typical line widths of the \ha\ emission are \( \sim 30 \) \kms,
  limiting temperatures in the ionized gas to \( <20,000 \) K. If
  radiation is the primary ionizing mechanism, the Lyman continuum
  flux required to sustain the most strongly emitting ionized regions
  is \( 1.2\times 10^{6} \) photons cm\( ^{-2} \) s\( ^{-1} \). There
  appears to be no local, stellar source capable of maintaining the
  ionization of the gas; however, the required ionizing flux is
  consistent with current models of the escape of Lyman continuum
  radiation from OB stars in the disk and of ionizing radiation
  produced by cooling supernova remnants.
\end{abstract}

\keywords{Galaxy: halo --- ISM: clouds --- ISM: individual (IVC-K) --- ISM:
structure --- ISM: atoms --- HII regions }

\section{Introduction\label{sec:intro}}

Optical emission lines are a new tool for the exploration of high-
and intermediate-velocity clouds (HVCs and IVCs), traditionally discovered
and studied in the 21 cm line of \hi\ or through absorption line observations.
The \ha\ intensity from such structures is typically a few tenths
of a Rayleigh (1 R\( =10^{6}/4\pi  \) photons cm\( ^{-2} \) s\( ^{-1} \)
sr\( ^{-1} \) \( =2.4\times 10^{-7} \) ergs cm\( ^{-2} \) s\( ^{-1} \)
sr\( ^{-1} \)) and has only recently been accessible through modern
detectors. With such faint emission, previous studies of emission
lines such as \ha, {[}\sii{]}, and {[}\nii{]} from HVCs and IVCs have
been limited to single pointed observations, although they have gleaned
some of the first information about the ionization, temperature, and
metallicity of these gas structures (\citealt{WW96}; \citealt*{TRH98};
\citealt{B-H+98}; \citealt{Wakker+99}). Since most of the known HVCs
and IVCs are at high latitudes and many are distant (\( d>500 \)
pc) structures, studies of their ionized component reveal information
not only about the clouds themselves but also about the environment
of the Galactic halo and the escape of radiation from the Galactic
plane.

With high sensitivity and velocity resolution, the recently completed
Wisconsin H-Alpha Mapper\footnote{%
More information about WHAM can be found at \url{http://www.astro.wisc.edu/wham/}.
} (WHAM) Northern Sky Survey is yielding an enormous amount of new
information about the ionized nature of IVCs. \citet{Wakker01} has
recently separated out the large (\( \Delta \ell \sim 40\arcdeg  \),
\( \Delta b\sim 40\arcdeg  \)) \hi\ intermediate-velocity Complex~K
(previously classified with Complex~C) centered near \( \ell =50\arcdeg  \),
\( b=+50\arcdeg  \), \( \vlsr \approx -60 \) to \( -80 \) \kms.
Here we present the first map of the ionized component of Complex~K.

\section{Observations\label{sec:observations}}

The WHAM Northern Sky Survey provides the first velocity-resolved
survey of \ha\ emission from the Galaxy. Each of the 37,300 pointings
in the northern survey captures a spectrum over \( \vlsr \approx -100 \)
to \( +100 \) \kms\ from a one-degree patch of sky. Since WHAM provides
12 \kms\ spectral resolution, we are able to cleanly remove the bright
and variable geocoronal \ha\ emission (\( I\sim 2-13 \) R) and numerous
fainter (\( I\sim 0.1-0.3 \) R) atmospheric lines (N. R. Hausen et
al., in preparation), allowing detections of faint Galactic emission
below 0.1 R. The survey achieves nearly full coverage of the sky with
beam spacings of \( \Delta b=0\fdg 85 \) and \( \Delta \ell =0\fdg 98/\cos b \)
above \( \delta >-30\arcdeg  \). More information about the instrument
and survey strategy can be found in \citet{TuftePhD}, \citet{HaffnerPhD},
and L. M. Haffner et al. (in preparation).

\section{Results\label{sec:results}}

Figure~\ref{fig:image} shows an image of the \ha\ emission from
WHAM and contours of 21 cm emission from the Leiden-Dwingeloo \hi\
survey \citep{HIAtlas} toward Complex K. Note that the image and contours
reflect a strict integration limit between \( -95 \) and \( -50 \)
\kms. As a result, the colorbar legend underestimates the total \ha\ line
intensity by about 20\% (\emph{e.g.} Figure~\ref{fig:spectrum1},
see below). Since the distance to Complex K is still relatively uncertain,
we have not applied an extinction correction to the data presented
here. However, since the total \( N_{HI}\, (-100<v_{LSR}<+100) \)
over the face of the cloud only varies from about 1 to \( 4\times 10^{20} \)
cm\( ^{-2} \), such a correction to the observed \ha\ flux would
be at most about 25\% if all the neutral material is between us and
Complex K.

The ionized emission is distributed diffusely over the neutral extent
of the complex; it does not appear to arise from numerous point sources
within the complex. The integrated \ha\ intensity in the region displayed
in Figure~\ref{fig:image} ranges from 0.5 R down to a conservative
detection limit of 0.1 R or EM \( \sim  \) 1.1 to 0.2 cm\( ^{-6} \)
pc (1 R \( \approx 2.25 \) cm\( ^{-6} \) pc for \( T=8000 \) K).
The extent of the \ha\ emission in the main part of the complex (\( \ell =30\arcdeg  \)
to \( 65\arcdeg  \), \( b=+30\arcdeg  \) to \( +70\arcdeg  \))
matches the \( N_{H}=7\times 10^{18} \) cm\( ^{-2} \) contour quite
closely; however, the peaks in each emission component do not correspond
as well. A detailed quantitative analysis of the \ha\ and \hi\ intensities
within each one-degree WHAM pointing reveals that there is little
correlation between the integrated intensities of the IVC components.
The linear correlation coefficient for pointings within the main part
of the complex is 0.4. Although the spatial distribution and velocity
profiles correspond extremely well, a simple relationship relating
the \ha\ intensity to the \hi\ column does not exist. In fact, for
many of the largest concentrations of \hi\ emission (\emph{e.g.,}
\( \ell =48\arcdeg ,\, b=+36\arcdeg  \); \( \ell =54\arcdeg ,\, b=+52\arcdeg  \);
\( \ell =45\arcdeg ,\, b=+44\arcdeg  \)), the brighter \ha\ emission
is consistently at higher longitudes than the regions of brightest
21 cm emission. We created \ha\ and \hi\ latitude profiles from the
images to quantify the extent of this offset. By examining the two
latitude profiles simultaneously and by computing correlation coefficients
between the profiles in their natural and offset positions, we find
that the \ha\ emission is consistently shifted or extended to higher
Galactic longitudes in nearly all latitude slices by a 2 to 4 degrees.

Figures~\ref{fig:spectrum1} and \ref{fig:spectrum2} present representative
\ha\ spectra (top panels) from Complex K with the one-degree WHAM
beam centered at \( \ell =57\fdg 02,\, b=+49\fdg 22 \) and \( \ell =48\fdg 03,\, b=+36\fdg 49 \),
respectively. Corresponding \hi\ 21 cm spectra centered at \( \ell =57\fdg 0,\, b=+49\fdg 0 \)
and \( \ell =48\fdg 0,\, b=+36\fdg 5 \) from the Leiden-Dwingeloo
\hi\ survey are displayed in the bottom panel of each figure. These
two sightlines sample some of the brighter emission from Complex K.
The dashed vertical lines denote the integration range used to construct
the \ha\ image and 21 cm contours in Figure~\ref{fig:image}. Since
hydrogen has a thermal width of \( \sim 22 \) \kms\ at \( T=10^{4} \)
K, we have chosen a restricted integration range to avoid contamination
in the image from the local emission component centered near \( \vlsr =0 \) \kms.
As noted above, the values in the color legend of Figure~\ref{fig:image}
should be increased by about 20\% to estimate the total intensity
of a typical Complex K emission line. 

The solid lines in the top panels of Figures~\ref{fig:spectrum1}
and \ref{fig:spectrum2} represent a Gaussian component fit convolved
with the WHAM instrument profile (approximately 12 \kms\ FWHM). We
also performed a multi-component fit on the \hi\ spectrum. Only the
IVC component is shown clearly in the bottom panels of Figures~\ref{fig:spectrum1}
and \ref{fig:spectrum2} since little effort was spent trying to accurately
model the bright, local 21 cm emission. Figure \ref{fig:spectrum2}
clearly shows at least two components in the intermediate-velocity
neutral gas. A single Gaussian fit to the \ha\ emission in this direction
results in a large (\( >40 \) \kms) width, atypical of diffuse gas;
a free-fit two Gaussian profile is not well defined due to the intrinsic
ionized gas line widths. Instead, we have chosen to fix the mean velocity
of the two components in the \ha\ fit to be the same as those clearly
defined in the \hi\ profile. The best parameters for our fits are
given in Table~\ref{tab:fit}. Channel maps and profile analyses show
some variation in the mean velocity of the \ha\ and \hi\ emission,
with values that fall roughly between \( -50 \) and \( -75 \) \kms.
There appears to be no systematic velocity gradients in the complex,
although the higher velocity \hi\ component at \( -75 \) \kms\ is
limited to the two regions of the complex centered at \( \ell =48\arcdeg ,\, b=+37\arcdeg  \)
and \( \ell =38\arcdeg ,\, b=+62\arcdeg  \).

\section{Discussion\label{sec:discussion}}

As reviewed by \citet{Wakker01}, little absorption line data directly
toward Complex K exists at this time, and thus we have little information
about its distance, only that it lies between about 0.3 and 7.7 kpc.
There is a firm upper limit since the cloud has been seen in absorption
\citep{dBS83,Bates+95,Shaw+96} toward M~13 (\( \ell =59\arcdeg ,\, b=+41\arcdeg  \)).
\citet{Carretta+00} have recently calculated the distance modulus
of M~13 to be 14.44 mag, which sets an upper limit of 7.7 kpc on
the distance to that portion of Complex K toward M~13. A lower limit
to the distance comes from the work of \citet{Bates+95} who included
three local stars in their absorption line study of stars in M~13.
Two of their targets, HD 151749 and HD 150462, have reliably measured
parallaxes \citep[\textit{HIPPARCOS}, ][]{Hipparcos}, which place
them at 150 and 175 pc, respectively. Their third target, HD 149802,
has an estimated distance of about 270 pc given that it and HD 150462
have both been classified as A0 but differ in V brightnesses by 0.9
mag. They did not detect Complex K absorption in either Ca II or Na
I toward any of these stars. 

If the entire complex is at a common distance then its angular extent,
from \( b=+30\arcdeg  \) to \( b=+70\arcdeg  \), corresponds to
a length of \( 0.68d \) kpc and a maximum vertical extent above the
midplane of \( 0.94d \) kpc, where \( d \) is the distance to the
complex in units of 1 kpc. If instead the complex is a column of gas
perpendicular to the Galactic plane at a constant \emph{radial} distance
(distance along the Galactic plane) from the sun, its length is \( 2.2r \)
kpc, where \( r \) is the radial distance of the complex from the
sun in units of 1 kpc. Gas participating in Galactic rotation toward
the center of the complex near \( \ell =55\arcdeg  \), \( b=+45\arcdeg  \)
would be expected to have a radial velocity of \( \vlsr =+10 \) \kms\
at 1 kpc up to a maximum (at the tangent point) of \( \vlsr =+30 \)
\kms\ at 6.5 kpc \citep{Clemens85}. As a result, the deviation velocity \citep{Wakker91}
for Complex K is approximately \( -100 \) \kms\ from that expected
by circular rotation.

With few observations of the ionized component of IVCs and HVCs, it
is difficult to discriminate among possible candidates for the source
of ionization. Since the clouds are moving at anomalous velocities,
processes such as ram-pressure heating \citep{WW96} and radiation
produced by a cloud-halo shocked interface have been explored \citep{TRH98}
in addition to photoionization by an ambient Lyman continuum flux
\citep{TRH98}. In the case of Complex K, the relatively diffuse \ha\ emission
over the extent of the neutral gas suggests the latter ionization
mechanism. However, at its lower distance limit (300 pc), the vertical
extent of the complex is only \( z=150 \) to \( z=300 \) pc, fully
within the WIM layer, and its anomalous velocity would likely cause
an interaction with the ambient medium. But unless Complex K is falling
nearly directly towards us, we find little evidence for an \ha\ enhancement
in the morphology due to cloud motion. In support of photoionization
by an external or ambient flux, note that while the total \hi\ column
densities presented in Table~\ref{tab:fit} for the IVC components
vary by a factor eight, the \ha\ intensities are relatively constant.
For the cloud as a whole, we find that the neutral gas displays much
higher intensity contrast than the ionized gas in Complex K (see also
Figures \ref{fig:spectrum1} and \ref{fig:spectrum2}). Furthermore,
as noted above in \S\ref{sec:results}, the point-to-point correspondence
of the total IVC intensity of the two emission lines is poor. These
observations are consistent with photoionizing the skin of a neutral
cloud. In this case, the cloud becomes self-shielding from ionizing
radiation and \( N_{H\, I} \) is independent from the \ha\ intensity,
which becomes mostly a function of incident Lyman continuum flux and
geometry.

If a cloud is ionized by an external photoionizing source, the incident
Lyman continuum flux on the cloud surface can be estimated \citep[see][]{TRH98}.
Ignoring the complexity of geometry but applying the maximum possible
extinction correction (25\%; see \S\ref{sec:results}) to the sample
ionized IVC component in Table~\ref{tab:fit}, we find the incident
flux of ionizing photons is \( \phi \approx 1.2\times 10^{6} \) \pcs\
for the brightest \ha\ regions in Complex K. With the small lower
limit on its distance, there is some chance that Complex K is a relatively
local feature that could be ionized by nearby sources. A search within
\( 16\arcdeg  \) of \( \ell =55\arcdeg  \), \( b=+45\arcdeg  \)
(the rough center of the brightest regions of the complex) reveals
no early spectral type objects with sufficient luminosity to ionize
the region, even at its minimum distance of \( \approx 300 \) pc.
As noted above, the ionized regions with higher \ha\ intensities appear
to be offset a few degrees from the larger column density neutral
regions. To test whether the ionization of the complex could be sustained
by a brighter, more distant halo source, we have explored a large
region at higher Galactic longitudes than Complex K (\( 80\arcdeg <\ell <180\arcdeg  \);
\( +30\arcdeg <b<+80\arcdeg  \)) for possible ionizing sources. In
this case, the only potential source appears to be the O9.5 V star
HD 93521 (\( \ell =183\arcdeg  \), \( b=+62\arcdeg  \), \( d=1.7 \)
kpc; \citealt{Irvine89}). Given the direction and range of distances
possible for Complex K, it is closest to the star when placed about
700 pc away (\( z\approx 500 \) pc); here they are separated by about
1.5 kpc. Using a value of \( \log Q_{0}=48.38 \), where \( Q_{0} \)
is the number of hydrogen ionizing photons per second emitted from
an O9.5 V star \citep*{VGS96} and taking into account \emph{only}
the geometrical dilution of the radiation (i.e. no intervening gas
or dust absorption), we find that the ionizing flux at Complex K from
HD 93521 is \( \phi \approx 8.4\times 10^{3} \) \pcs, far short of
that needed to maintain the ionization of the complex.

More likely, a scenario similar to those used to explain the ionization
of the Warm Ionized Medium \citep*{MC93,DS94,DSF00,SMH00} is also
providing the ionizing flux for Complex K. \citet{DSF00} cite upward
halo ionizing fluxes for their various models of the propagation of
OB association radiation through the surrounding superbubbles. Their
range of calculated values, \( \phi \sim 1 \)--\( 2\times 10^{6} \)
\pcs, agree well with what is required to maintain the ionization
of Complex K. \citet{B-HM99} have produced a model of the propagation
of the ionizing flux in the Galactic halo and the resulting \ha\ emission
from halo clouds. We find that our observed \ha\ intensity is consistent
with that expected from their model in the direction of Complex K
for distances \( <8 \) kpc from the sun (0.15--0.30 R; J. Bland-Hawthorn
2000, private communication). Cooling hot gas generated in supernova
remnants has also been shown by \citet{SMH00} to provide a mean value
of \( \phi =5.4\times 10^{5} \) \pcs\ to the halo ionizing flux,
nearly half that required for the brightest emitting regions in the
complex and sufficient alone to sustain the faintest.

With WHAM, we have finally begun to explore the structure of the ionized
component of IVCs and HVCs with the \ha\ line in the same manner as
has been done for the neutral gas with the 21 cm line. Furthermore,
maps in other optical emission lines such as {[}\sii{]} and {[}\nii{]}
will provide additional information about the temperature, ionization,
and metallicity of the ionized gas \citep{HRT99,Wakker+99}. These
future observations have the potential to provide insight not only
about the nature of these high latitude structures but also about
the ambient Galactic halo gas and radiation field.

\acknowledgements{We would like to thank Bart Wakker for the first
  \hi-\ha\ comparisons in this region, Joss Bland-Hawthorn for
  providing model results along the Complex K sight line, and Bob
  Benjamin for continuing halo consultation and Galactic rotation
  code. Source searches were conducted through the SIMBAD database
  operated at CDS, Strasbourg, France. WHAM is supported by the
  National Science Foundation through grant AST 96-19424.}


\clearpage

\begin{deluxetable}{clr@{$\pm$}lr@{$\pm$}lr@{$\pm$}ll}

\tablewidth{0pt}

\tablecaption{Spectral Fit Parameters\label{tab:fit}}

\tablehead{

\colhead{Direction} & \colhead{Component} &

\multicolumn{2}{c}{Mean} & 

\multicolumn{2}{c}{FWHM} & 

\multicolumn{3}{l}{Intensity} \\

\colhead{$(\ell, b)$} & \colhead{} &

\multicolumn{2}{c}{[\kms]} & 

\multicolumn{2}{c}{[\kms]} & 

\multicolumn{3}{c}{} 

}

\startdata

$(57\fdg02, +49\fdg22)$ & \ha\ Local & $-4.1$ & 0.8 & 27.0 & 2.5 & 0.50 & 0.03 & R \\

$(57\fdg02, +49\fdg22)$ & \ha\ IVC & $-64.0$ & 0.8 & 29.0 & 2.3 & 0.48 & 0.02 & R \\

$(57\fdg00, +49\fdg00)$ & \hi\ IVC & $-60.4$ & 1.0 & 22.0 & 2.5 & 8.9 & 0.9 & $10^{18}$ cm$^{-2}$ \\

\\

$(48\fdg03, +36\fdg49)$ & \ha\ Local & $-4.5$ & 0.5 & 23.9 & 1.4 & 0.78 & 0.02 & R \\

$(48\fdg03, +36\fdg49)$ & \ha\ IVC & \multicolumn{2}{c}{$-47$\tablenotemark{a}} & 22.2 & 10.0 & 0.15 & 0.03 & R \\

$(48\fdg03, +36\fdg49)$ & \ha\ IVC & \multicolumn{2}{c}{$-73$\tablenotemark{a}} & 24.7 & 4.6 & 0.27 & 0.03 & R \\

$(48\fdg00, +36\fdg50)$ & \hi\ IVC & $-46.8$ & 1.1 & 19.1 & 2.2 & 1.9 & 0.3 & $10^{19}$ cm$^{-2}$ \\

$(48\fdg00, +36\fdg50)$ & \hi\ IVC & $-72.8$ & 0.6 & 25.7 & 1.1 & 5.3 & 0.3 & $10^{19}$ cm$^{-2}$ \\

\enddata

\tablenotetext{a}{For the fit in this direction, the mean velocities of the ionized IVC components have been fixed to be the same as those derived from the fit to the neutral gas. See \S\ref{sec:results} for full details.}

\end{deluxetable}

\clearpage

\figcaption[fig1.ps]{The ionized and neutral components of Complex K.
  Integrated \ha\ emission from \protect\( \vlsr =-95\protect \) to
  \protect\( -50\protect \) kms is presented as the color image. \hi\ 
  column densities (integrated over the same velocity range) of 7, 14,
  and \protect\( 21\times 10^{18}\protect \) cm\protect\(
  ^{-2}\protect \) from Hartmann \& Burton (1997) are displayed as
  light blue contours.\label{fig:image}}

\figcaption[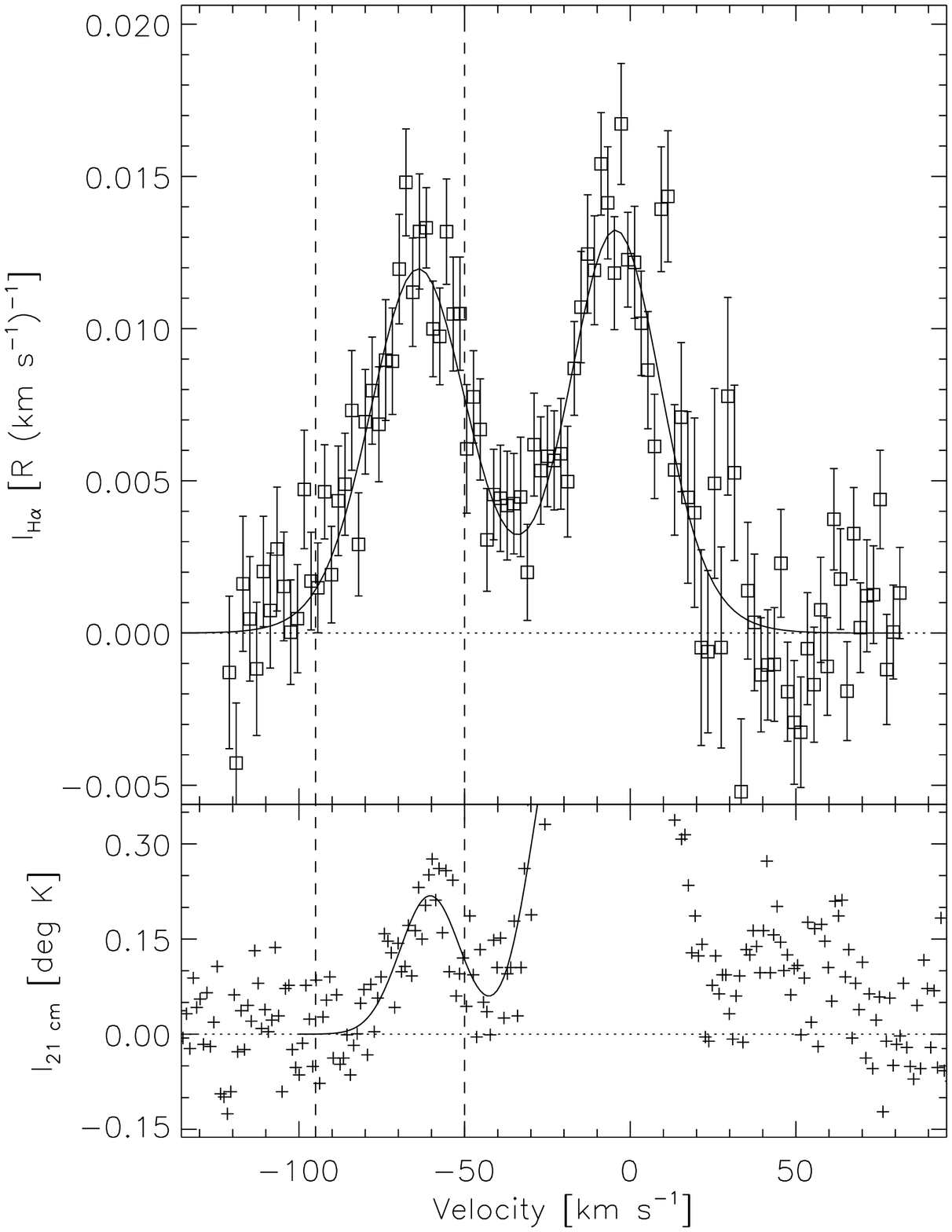]{(\emph{top}) WHAM \ha\ spectrum of the one-degree
  portion of the sky centered at \protect\( \ell =57\fdg 02,\,
  b=+49\fdg 22\protect \).  WHAM data is plotted as squares with one
  sigma error bars. A 3.7 R geocoronal line centered near \protect\(
  \vlsr =+30\protect \) \kms\ has been removed from the spectrum. A
  simple two component fit to the Galactic emission is shown as the
  solid line (see text). (\emph{bottom)} \hi\ 21 cm emission spectrum
  toward \protect\( \ell =57\fdg 0,\, b=+49\fdg 0\protect \) from the
  Leiden-Dwingeloo \hi\ survey. The local emission peaks off the
  displayed scale near 8.2 K. The solid line denotes the IVC portion
  of a multi-component fit (see text). In both panels, the dashed
  lines denote the integration range used for the image in
  Figure~\ref{fig:image}.
\label{fig:spectrum1}}

\figcaption[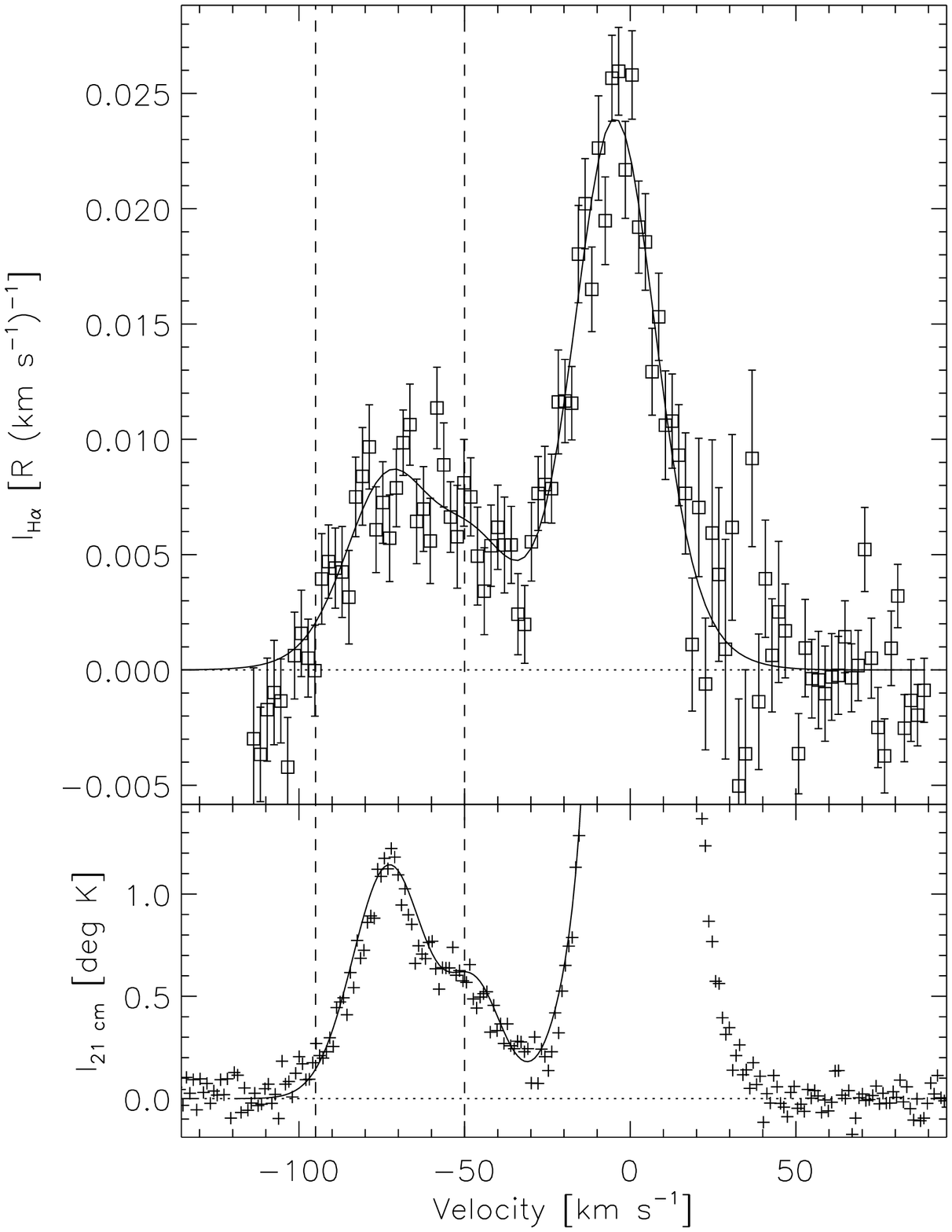]{Same as Figure~\ref{fig:spectrum1} except toward
  (\emph{top}) \protect\( \ell =48\fdg 03,\, b=+36\fdg 49\protect \)
  for \ha\ and (\emph{bottom}) \protect\( \ell =48\fdg 0,\, b=+36\fdg
  5\protect \) for \hi. The local \hi\ emission peaks off the
  displayed scale near 15.8 K. \label{fig:spectrum2}}

\begin{figure}[htbp]
  \begin{center}
    \includegraphics[width=\textwidth]{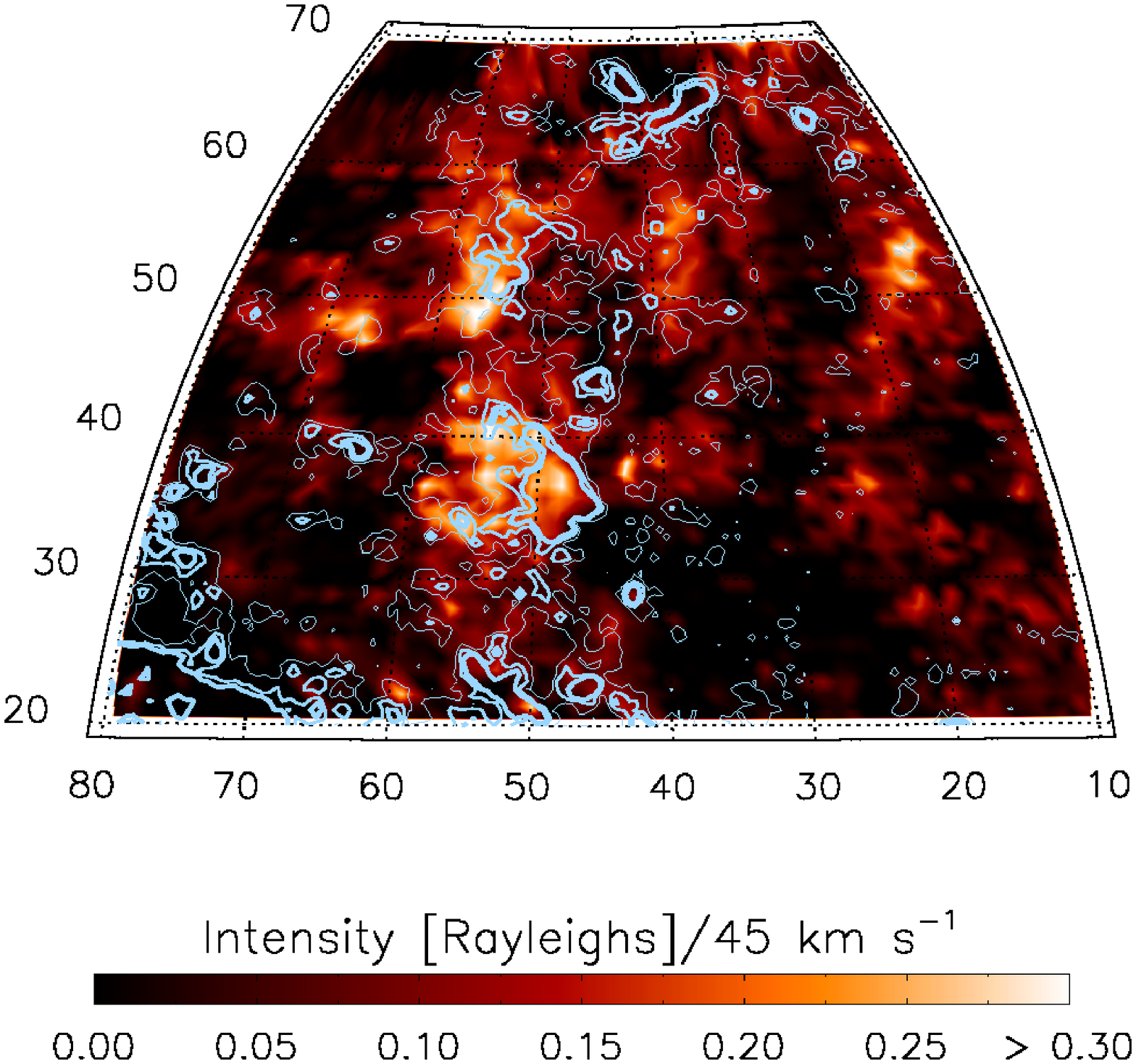}
  \end{center}
\end{figure}

\begin{figure}[htbp]
  \begin{center}
    \includegraphics[width=\textwidth]{fig2.ps}
  \end{center}
\end{figure}

\begin{figure}[htbp]
  \begin{center}
    \includegraphics[width=\textwidth]{fig3.ps}
  \end{center}
\end{figure}

\end{document}